\documentclass{article}
\pdfoutput=1
\usepackage{amsthm}
\usepackage{graphicx}
\usepackage{float}
\usepackage{color}
\usepackage{amsmath}
\usepackage{amssymb}
\usepackage{multirow}
\usepackage{multicol}
\usepackage{authblk}
\usepackage{array}
\usepackage{epstopdf}
\usepackage{mathtools}

\author[1]{Glenn Young}
\author[1]{Pengcheng Xiao}
\author[2]{Kenneth Newcomb}
\author[2]{Edwin Michael}

\affil[1]{Department of Mathematics, Kennesaw State University, Marietta, GA, USA}
\affil[2]{Center for Global Health Infectious Disease Research, University of South Florida, Tampa, FL, USA }
\begin{document}

\title{Interplay between COVID-19 vaccines and social measures for ending the SARS-CoV-2 pandemic}
\date{\today}

\maketitle

\abstract{
The development and authorization of COVID-19 vaccines has provided the clearest path forward to eliminate community spread hence end the ongoing SARS-CoV-2 pandemic. However, the limited pace at which the vaccine can be administered motivates the question, to what extent must we continue to adhere to social intervention measures such as mask wearing and social distancing? To address this question, we develop a mathematical model of COVID-19 spread incorporating both vaccine dynamics and socio- epidemiological parameters. We use this model to study two important measures of disease control and eradication, the effective reproductive number $R_t$ and the peak intensive care unit (ICU) caseload, over three key parameters: social measure adherence, vaccination rate, and vaccination coverage. 
Our results suggest that, due to the slow pace of vaccine administration, social measures must be maintained by a large proportion of the population until a sufficient proportion of the population becomes vaccinated for the pandemic to be eradicated. By contrast, with reduced adherence to social measures, hospital ICU cases will greatly exceed capacity, resulting in increased avoidable loss of life. These findings highlight the complex interplays involved between vaccination and social protective measures, and indicate the practical importance of continuing with extent social measures while vaccines are scaled up to allow the development of the herd immunity needed to end or control SARS-CoV-2 sustainably.}

\section{Introduction}
 The advent of COVID-19 vaccines and mass vaccinations of populations have led to widespread public expectation that we may be able to end the ongoing SARS-CoV-2 pandemic in some economically advanced countries by as early as the of beginning 2022  \cite{Randall2021}. While the pace at which these vaccines have been developed and authorized by governments for population-wide usage has been unprecedented \cite{Lurie2020} reflecting the desire to fast track the ending or control of the pandemic given the socio-economic costs of protracted non-pharmaceutical interventions (NPIs), such as cyclical lockdowns and social distancing measures \cite{Ebrahim2020}, it is also clear that several features of the current vaccines and vaccination strategies for achieving this goal remain unresolved \cite{Buckner2020,Lipsitch2020}. 

First, it is important to consider that vaccines serve two major purposes: to protect the individual from contracting the disease and to stop the transmission of community infection. While the initial vaccine trial data for the 3 major vaccines approved for use thus far in developed world settings, viz. Pfizer-Biontech, Moderna, and AstraZeneca-Oxford, indicate that these could induce very high levels of protection (70-90\%) against symptomatic disease \cite{Terry2021}, more recent data with regard to the AstraZeneca vaccine suggests that vaccination may also reduce community transmission of the virus significantly \cite{Wise2021}. These data suggest that both disease outcomes and transmission could be significantly reduced in communities as a result of mass deployments of these vaccines. A key factor, however, is that in both cases current vaccines are not 100\% protective. Second, it is apparent that the number of vaccines initially available and the logistical challenges connected with their delivery will hamper the rapid vaccination of a population, which will prolong the time to eradicating the disease through vaccination in populations {\cite{OurWorld2021}. An added challenge is the reduced effectiveness of these vaccines as currently formulated against newly emerging virus variants \cite{Davies2020}.

A third important factor is to consider the epidemiological and social contexts in which vaccinations will take place. This is important because many populations undergoing vaccinations will have already experienced one or more waves of COVID-19, as a result of which some level of natural immunity to SARS-CoV-2 will likely to be in operation in these communities. Such pre-existing immunity could indicate that the vaccine coverages to end the pandemic (reduce the prevailing effective reproduction number, $R_t$ to below 1 sustainably) need not be too high even if vaccine effectiveness is not perfect \cite{Iboi2020}, increasing the prognosis for using the current vaccines for ending the pandemic. On the other hand, these populations are also currently experiencing various levels of NPIs \cite{Ebrahim2020}. Such social mitigation or containment measures, while protecting the susceptible fraction from infection will act also to depress the development of natural immunity in the population. These outcomes suggest that there may be complex interactions between the two interventions, a better understanding of which will be crucial for determining how best to optimally deploy these tools for controlling or ending the pandemic. Further, investigating such interactions will also be important to fully understanding the implications of relaxing these NPIs as vaccinations roll out \cite{Iboi2020,Jentsch2020}.

Here, we extend our existing data-driven socio-epidemiological SEIR-based COVID-19 mathematical model \cite{Newcomb2020} by incorporating imperfect vaccination dynamics in order to undertake a theoretical investigation of this topic. To enhance realism we use the basic model calibrated to infection data on the course of the pandemic in the Tampa Bay region, and use the resulting model to investigate the interplay between vaccination and social protective measures for effective control or elimination of the pandemic. In particular, we explore the dynamical implications of imperfect vaccine effectiveness, vaccine rollout rates, and coverages of vaccination and social measures on the eradication of the disease in the community, and the effects these will have for virus transmission and critical care requirements. We also inspect the effect that the current vaccination roll out will have on the extent to which a population may relax social measures to return to normalcy.

\section{Methods}

\subsection{Model formulation}

Here we develop an extended SEIR model to assess population-level disease dynamics of COVID transmission. We consider a population of fixed size that we divide into eleven interacting subpopulations: susceptible without access to a vaccine ($S$), vaccinated susceptible ($V$), unvaccinated susceptible (with access to a vaccine) ($X$), exposed but not intectious ($E$), asymptomatic infected ($I_a$), pre-symptomatic infected ($I_p$), mildly symptomatic infected ($I_m$), hospitalized infected ($I_h$), critical care infected ($I_c$), recovered ($R$), and died ($D$). Since the population size is fixed, we can impose the condition that $S+X+V+E+I_a+I_p+I_m+I_h+I_c+R+D=1,$ and each population can therefore be interpreted as a proportion of the total population.

Our model is visualized as a diagram in Figure \ref{fig:modelDiag}. Importantly, we assume that asymptomatic, pre-symptomatic, and mildly symptomatic individuals transmit the disease at the same rate $\beta$, and that these are the only sources of transmission. The transmission rate is reduced due to social measures (face masks, social distancing, self quarantine, etc) by a factor of $1-ac$ among all infectious individuals, where $c$ is the efficacy of social measures and $a$ is the proportion of the population adhering to social measures.

\begin{figure}[h!]
{\centering
\includegraphics[width=0.9\textwidth]{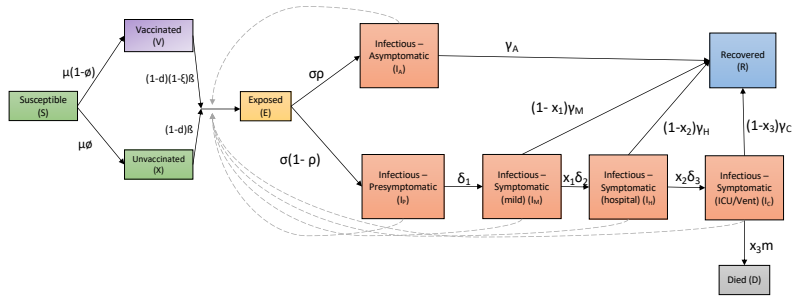}

}
\caption{Model diagram}
\label{fig:modelDiag}
\end{figure}

The vaccine becomes available to the population at rate $\mu$, and individuals can choose whether or not to get the vaccine. Specifically, proportion $\phi$ of the susceptible population $S$ enters the vaccinated population $V$ at rate $\mu$, while proportion  $1-\phi$ of the susceptible population enters the unvaccinated class $X$ at the same rate. This parameter $\phi$ allows us to analyze the effects of vaccine coverage, specifically as a consequence of individuals who are unwilling to receive the vaccine \cite{Young2015b}. Vaccinated individuals are infected at a rate reduced by $1-\xi$, where $\xi$ is the efficacy of the vaccine, while the unvaccinated class $X$ has the same dynamics as the susceptible without access to a vaccination class $S$.

With these assumptions, our model is given by the following system of eleven ordinary differential equations:

\begin{equation}\label{eq:model}
\begin{aligned}
S'&=-(1-d)\beta S(I_a+I_p+I_m)-\mu S\\
X'&=\mu(1-\phi)S-(1-d)\beta X(I_a+I_p+I_m)\\
V'&=\mu\phi-(1-\xi)(1-d)\beta V (I_a+I_p+I_m)\\
E'&=(1-d)\beta (I_a+I_p+I_m)\left[S+X+(1-\xi) V\right]-\sigma E\\
I_a'&=\sigma\rho E-\gamma_a I_a\\
I_p'&=\sigma(1-\rho)E-\delta_1I_p\\
I_m'&=\delta_1I_p-x_1\delta_2I_m-(1-x_1)\gamma_mI_m\\
I_h'&=x_1\delta_2I_m-x_2\delta_3I_h-(1-x_2)\gamma_hI_h\\
I_c'&=x_2\delta_3I_h-(1-x_3)\gamma_cI_c-x_3mI_c\\
R'&=\gamma_aI_a+(1-x_1)\gamma_mI_m+(1-x_2)\gamma_hI_h+(1-x_3)\gamma_cI_c\\
D'&=x_3mI_c
\end{aligned}
\end{equation}
The state variables and parameters are defined in Tables \ref{tab:vars} and \ref{tab:params}, respectively. The parameters are estimated by the methods described in Appendix \ref{sec:params}.

\begin{table}[h!]
{\centering 
\begin{tabular}{  c  l  }
\hline
\multicolumn{2}{c}{State Variable} \\
\hline\hline \noalign{\smallskip}
$S$ & Susceptible population without access to a vaccine\\
$X$ & Susceptible, unvaccinated population with access to a vaccine\\
$V$ & Susceptible, vaccinated population\\
$E$ & Exposed population (pre-infectious) \\
$I_a$ & Asymptomatic, infectious population\\ 
$I_p$ & Pre-symptomatic, infectious population\\
$I_m$ & Symptomatic, infectious population without need of hospitalization\\
$I_h$ & Hospitalized infected population\\
$I_c$ & Infected population in intensive care unit (ICU)\\
$R$ & Recovered population\\
$D$ & Deceased population
\end{tabular}

}
\caption{ 
State variables. Each variable should be interpreted as a proportion of the total population.}
\label{tab:vars}
\end{table}

\subsection{Parameter estimation}\label{sec:params}

We used a Monte Carlo-based Bayesian Melding approach to parameterize the base model using case notification, death, and mobility data reported for the Tampa Bay region for the period between March 10th and August 24th 2020. (details of methods provided in Newcomb et al. 2020). Briefly, all social and epidemiological model parameters that could not be fixed at initiation were sequentially updated using 10-day blocks or segments of data between the above estimated period from their initial prior values using this procedure, and the final model thus estimated was used for the simulations carried out in this paper. Confirmed case data for the four counties comprising Tampa Bay, viz. Hillsborough, Pasco, Pinelles and Polk, were obtained from the Johns Hopkins University Coronavirus Resource Center (Dong et al. 2020). Mobility data serve as an estimate for population mixing and the fraction of the population under restricted movement in the counties concerned were obtained from the location data firm, Unacast (https://www.unacast.com/covid19/social-distancing-scoreboard).

\begin{table}[h!]
{\centering 
\begin{tabular}{  c  l  c }
\hline
\multicolumn{2}{c}{Parameter} & Value \\
\hline\hline \noalign{\smallskip}
$\beta$ & Transmission rate & 0.76\\
$a$ & Proportion of population adhering to social measures & varies\\
$c$ & Reduction in transmission due to social measures & 0.85\\
$\phi$ & Vaccine coverage & varies\\
$\xi$ & Vaccine efficacy & 0.9\\ 
$\mu$ & Rate of vaccination & 0.02\\
$\sigma$ & Rate of symptom onset among exposed & 0.32\\
$\rho$ & Proportion of expose who never develop symptoms & 0.37\\
$\gamma_a$ & Rate of recovery among $I_a$ & 0.23 \\
$\gamma_m$ & Rate of recovery among $I_m$ & 0.23\\
$\gamma_h$ & Rate of recovery among $I_h$ & 0.22\\
$\gamma_c$ & Rate of recovery among $I_c$ & 0.23 \\
$\delta_1$ & Rate of progressing from $I_p$ to $I_m$ & 0.69\\
$\delta_2$ & Rate of progressing from $I_m$ to $I_h$ & 0.15\\
$\delta_3$ & Rate of progressing from $I_h$ to $I_c$ & 0.55\\
$x_1$ & Proportion of mild cases that progress to hospital & 0.18\\
$x_2$ & Proportion of hospital cases that process to ICU & 0.25\\
$x_3$ & Proportion of ICU cases that  die & 0.51\\
$m$ & Mortality rate among ICU patients & 0.37
\end{tabular}

}
\caption{ 
Parameters. The value for each parameter is used throughout this work unless stated otherwise. All rate parameters are measured in units of (day)$^{-1}$.}
\label{tab:params}
\end{table}

\subsection{Effective reproduction number}

The effective reproductive number $R_t$ quantifies the average number of secondary infections caused by each new infection. 
We can calculate $R_t$ using the next generation matrix method developed in \cite{vanDenDriessche2002}. The next-generation matrices for system \eqref{eq:model} are

$$F=\begin{bmatrix}
0 & \lambda & \lambda  & \lambda\\
0 & 0 & 0 & 0 \\
0 & 0 & 0 & 0 \\
0 & 0 & 0 & 0 
\end{bmatrix}$$
and

$$V=\begin{bmatrix}
-\sigma & 0 & 0  & 0 \\
-\rho \sigma & \gamma_A & 0 & 0 \\
-(1-p)\sigma & 0 & \delta_1 & 0 \\
0 & 0 & -\delta_1 & x_1\delta_1+(1-x_1)\gamma_M 
\end{bmatrix},$$
where  $$\lambda =(1-ac)\beta \left(S+X+(1-\xi)V\right)$$  is the force of infection.

Then the effective reproduction number $R_t$ is exactly the spectral radius of $FV^{-1}$:
\begin{equation}\label{eq:Rt}
R_t=\lambda\left(\frac{1-\rho}{\delta_1}+\frac{1-\rho}{(1-x_1)\gamma_M+x_1\delta_2}+\frac{\rho}{\gamma_A}\right).
\end{equation}
We use this formula to assess the viability of vaccination programs on disease eradication in the following sections.

\section{Results} 

\subsection{Disease control dynamics}

We begin by determining the extent to which a population can relax social measures once a vaccine becomes available and still control the virus. Here, we use simulations of the number of individuals requiring intensive care as a measure of control, and thus consequently, we consider the relationship between vaccine coverage $\phi$ and social measure compliance $a$ on ICU hospitalization to address this question. Figure \ref{fig:timeSeries} shows ICU hospitalizations over time under various vaccine coverages and social measures. The red curve shows ICU cases without any vaccine or social measures, serving as a baseline for comparison. The black curve shows these cases with some vaccination and no social measures; the green curve shows ICU cases with strong social measures and no vaccine; the blue curve shows ICU cases with both vaccine coverage and social measures. Of course, greater vaccine coverage and stronger social measures result in fewer cases; however, social measures reduce the incidence of cases requiring intensive care much more than vaccine coverage. This is due to the slow vaccination rate $\mu$, as we show in Section \ref{sec:time}.

\begin{figure}[h!]
{\centering
\includegraphics[width=0.9\textwidth]{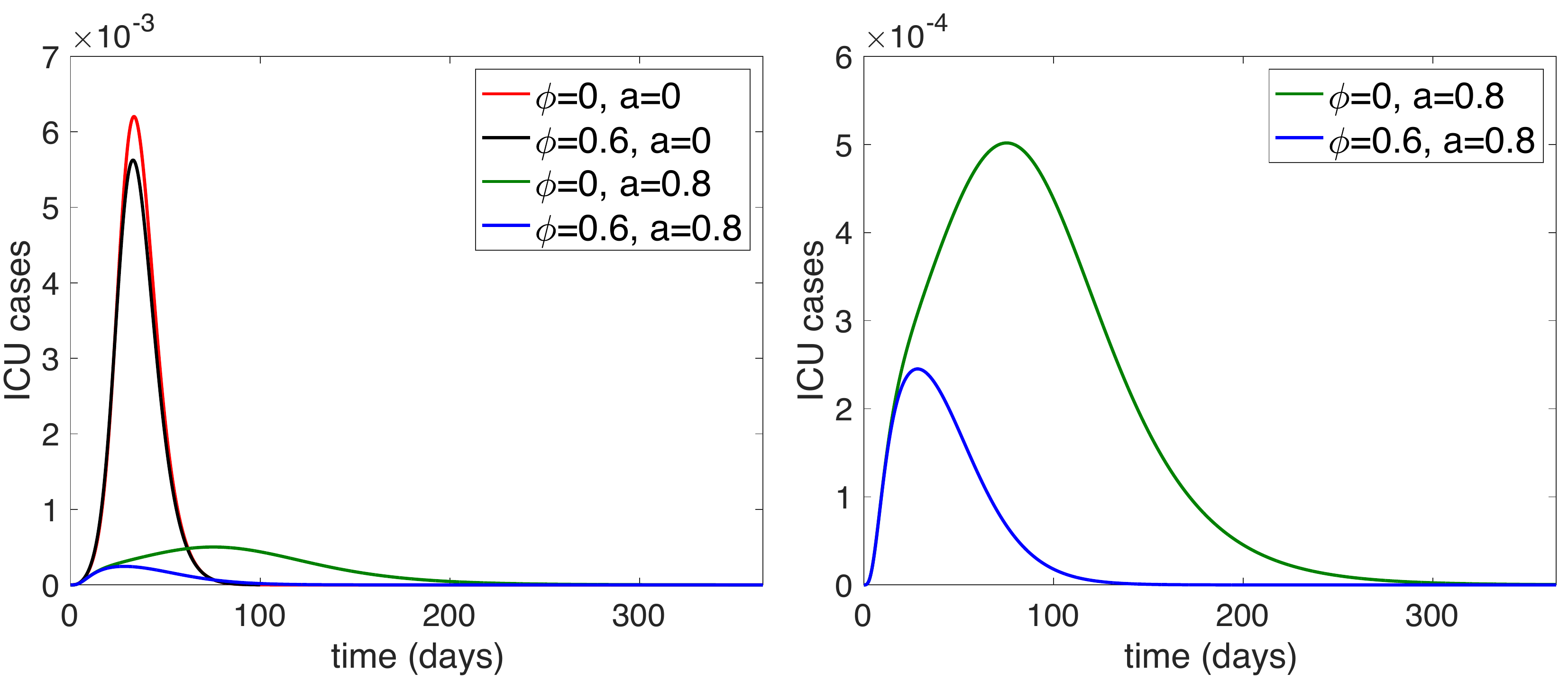}

}
\caption{ICU cases under various vaccination coverages and social measures. The vaccination rate and efficacy are $\mu=0.02$ and $\xi=0.94$, respectively. The right panel shows the same blue and green curves from the left panel, but zoomed in. All simulations performed using XPPAUT \cite{Bard2002}.}
\label{fig:timeSeries}
\end{figure}

\subsection{Conditions for eradication}

We use the effective reproduction number $R_t$ to study strategies by which the virus can be eradicated. Importantly, $R_t=R_t(t)$ is a function of time due to its dependence on  $S$, $X$, and $V$: as the susceptible populations decrease, so does $R_t$. We consider the temporal dynamics of $R_t$ in the following section. Here, we consider the idealized case in which proportion $\phi$ of the population is \textit{instantaneously} vaccinated, $V=\phi S_r$ and $X=(1-\phi)S_r$, where $S_r$ is a new parameter representing the remaining susceptible proportion of the population at the time the vaccine is administered (that is, $1-S_r$ is the proportion of the population who are currently infected, recovered, or deceased). In this case, $\lambda$ can be written

\begin{equation*}
\lambda=(1-ac)\beta S_r \left(1-\xi\phi \right),
\end{equation*}
and so

\begin{equation}\label{eq:Rt}
R_t=(1-ac)\beta S_r \left(1-\xi\phi \right)\left(\frac{1-\rho}{\delta_1}+\frac{1-\rho}{(1-x_1)\gamma_M+x_1\delta_2}+\frac{\rho}{\gamma_A}\right).
\end{equation}

The ideal goal in eradicating the disease is to permanently reduce the effective reproductive number below the threshold $R_t=1$. Figure \ref{fig:Rt1} shows curves in parameter space satisfying $R_t=1$ with $\xi=0.94$. On the left, curves are plotted over vaccine coverage $\phi$ and social measure compliance $a$ and remaining susceptible population $S_r=1$ (red curve) and $S_r=0.8$ (blue curve). In order to achieve $R_t<1$, $(\phi,a)$ pairs must lie above the curve. Note that disease eradication without social measures ($a=0$) would require more than $80\%$ of the population to receive a vaccine that is $94\%$ effective in a naive population ($S_r=1$); a population in which 20\% have already been exposed would require approximately 75\% to receive a vaccine in order to eradicate the disease without social measures. On the right, we show curves defined by $R_t=1$ over vaccine coverage $\phi$ and remaining susceptible population $S_r$ for varied social measures. For each fixed value of $a$, $(\phi,S_r)$ pairs must be below the curve to drive $R_t<1$. Without social measures in place (red curve), approximately 75\% of the population would have to be exposed to the virus before the disease is eradicated without a vaccine. In general, the larger the remaining susceptible population, the higher the vaccination coverage required to eradicate the disease. Importantly, the vaccine coverage threshold necessary to drive $R_t<1$ decreases with increased social measures. 

\begin{figure}[h!]
{\centering
\includegraphics[width=0.9\textwidth]{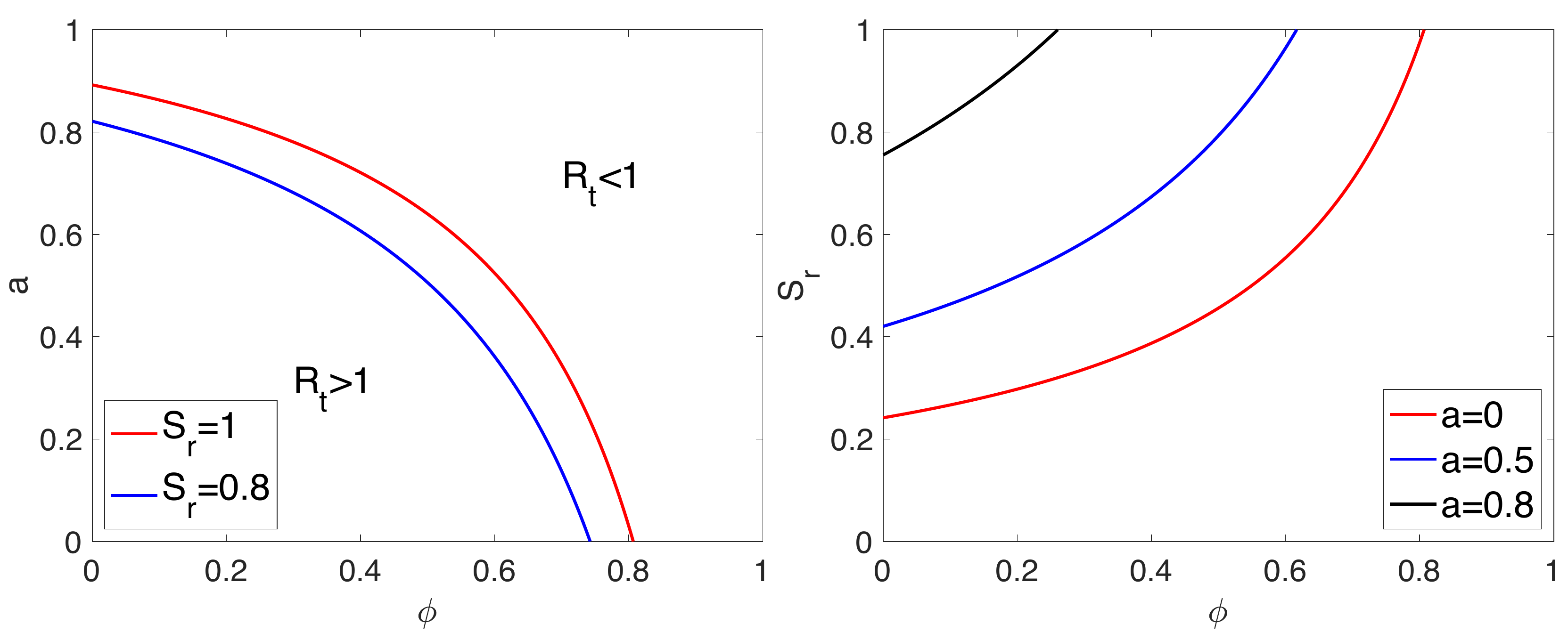} 

}
\caption{Vaccination and social measure thresholds to achieve $R_t<1$. Each curve defines $R_t=1$ over varied parameters with $\xi=0.94$. (Left)  $R_t=1$ plotted over vaccine coverage $\phi$ and social measure compliance $a$. The region above each curve represents parameter pairs for which $R_t<1$. (Right) $R_t=1$ plotted over vaccine coverage $\phi$ and remaining susceptible population $S_r$ with no, medium, and high proportions of social measure compliance. The region below each curve represents parameter pairs for which $R_t<1$.}
\label{fig:Rt1}
\end{figure}

\subsection{Time to eradication: interactions between vaccination coverage and rate and social measures}\label{sec:time}
Vaccine rollout will not be instantaneous; it will likely take months to vaccinate a majority of the population. We therefore must consider the simultaneous effects of infection dynamics and slow vaccination rates on disease eradication, which requires considering the effective reproduction number as a function of time:
\begin{equation}\label{eq:RtTime}
R_t(t)=(1-ac)\beta \left(S(t)+X(t)+(1-\xi)V(t)\right)\left(\frac{1-\rho}{\delta_1}+\frac{1-\rho}{(1-x_1)\gamma_M+x_1\delta_2}+\frac{\rho}{\gamma_A}\right).
\end{equation}
As the three susceptible populations $S$, $X$, and $V$ change over time via infection and vaccination dynamics, $R_t(t)$ strictly decreases. We denote the time at which $R_t(t)$ decreases below 1 by $t_h$; that is, $R_t(t_h)=1$. Time $t_h$ marks the beginning of the end of community spread, and we therefore refer to $t_h$ as the time to eradication. The left panel of Figure \ref{fig:Rt1phi} shows $t_h$ as a function of vaccination coverage $\phi$ for varied social measure compliance $a$ with vaccination rate $\mu=0.02$. The time to eradication remains nearly constant over all $\phi$ for low social measure compliance ($a=0,0.5$; red and blue curves): this is due to the slow vaccination rate $\mu$, and the comparably fast infection rate due to low social measure compliance. That is, for sufficiently small $a$, the virus spreads quickly through the population, infecting the susceptible population $S$ much more quickly than the susceptibles become vaccinated. Thus, eradication is achieved primarily through infection, rather than through vaccination. For high social measure compliance ($a=0.8$; black curve), $t_h$ is large for small $\phi$. This is because, for large $a$, the infection dynamics are slowed down, but because $\phi$ is small, the vaccination rate is also slow. The two mechanisms by which eradication is achieved (infection and vaccination) are therefore both slow, and so $t_h$ is large. As $\phi$ increases, however, $t_h$ decreases dramatically. The effective reproduction number $R_t$ decreases as $a$ increases, and thus achieving $R_t=1$ requires less vaccination and infection for large $a$. Thus, despite the slow vaccination rate $\mu$, there is a critical $\phi$ value past which the susceptible population becomes vaccination quickly enough so that $R_t$ decreases below 1 due primarily to vaccine administration. In other words, the vaccination timescale overtakes the infection rate timescale for sufficiently large $\phi$.

The right panel of Figure \ref{fig:Rt1phi} shows the maximum ICU case load (that is, the peaks of the curves in Figure \ref{fig:timeSeries}) as a function of vaccination coverage for varied $a$. Unsurprisingly, as $\phi$ increases, the maximum ICU load decreases for each $a$. However, the peak ICU load remains comparatively high for $a=0$ and $a=0.5$ (red and blue curves, respectively) compared with $a=0.8$ (black curve). For this latter case, the peak ICU load decreases dramatically from $\phi=0$ until around $\phi=0.4$, then remains low for all larger $\phi$. ICU capacity in most states is between $10^{-4}$ and $3\times 10^{-4}$: peak ICU cases only remain below this threshold for high social measure compliance and vaccination coverage.

\begin{figure}[h!]
{\centering
\includegraphics[width=0.9\textwidth]{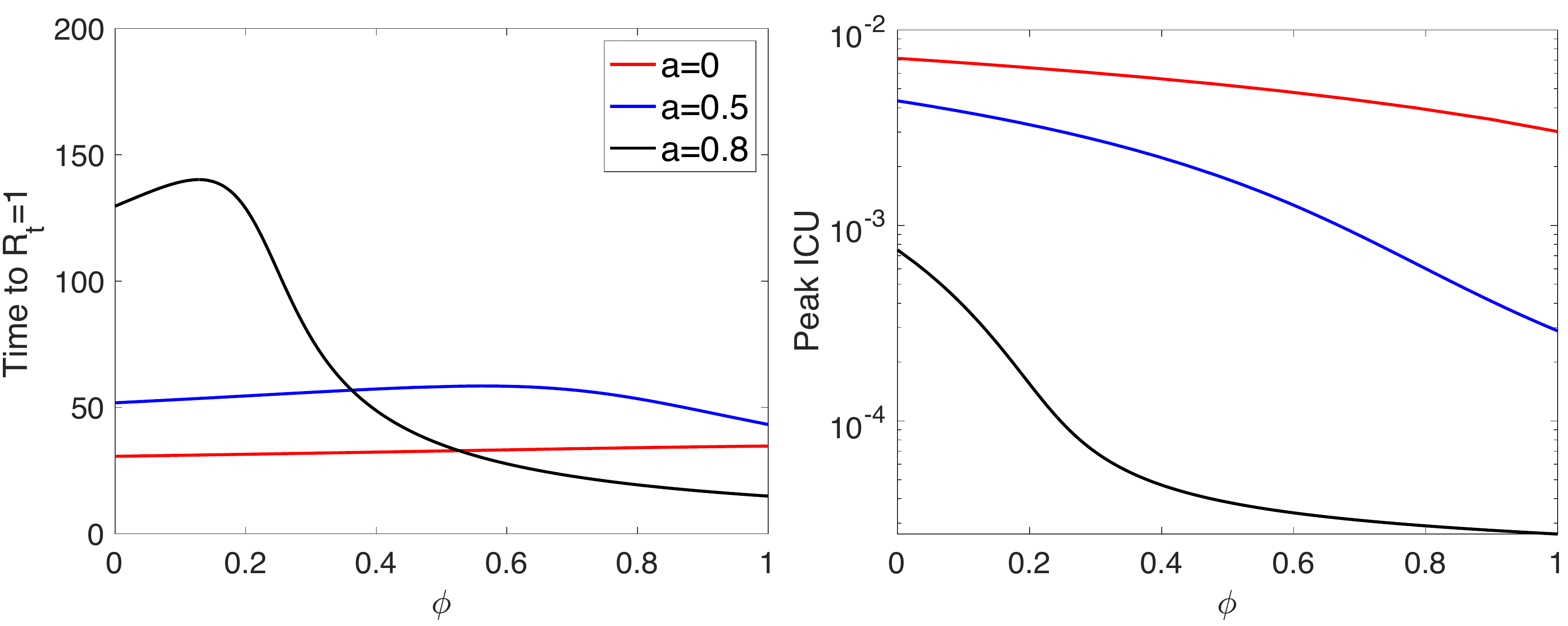}

}
\caption{Controlling the pandemic with a vaccine depends on broad coverage and sustained social measure compliance. (Left) Time until $R_t<1$ over $\phi$. (Right) Peak ICU cases over $\phi$ (log scale). In both panels, $\mu=0.02$, and the red curve corresponds to no social measures ($a=0$), the blue curve corresponds to moderate social measure compliance ($a=0.5$), and and the black curve corresponds to high compliance ($a=0.8$). Details on how these figures were made are in Appendix \ref{sec:BVPs}.}
\label{fig:Rt1phi}
\end{figure}


Figure \ref{fig:Rt1phi} suggests that sustained social measures help to eradicate the disease more efficiently than vaccination programs. This at least in part due to a relatively slow vaccination rate: $\mu=0.02$. We now investigate the influence of vaccination rate $\mu$ on the time to eradication and on peak ICU cases. The left panel of Figure \ref{fig:Rt1mu} shows the time until the disease is eradicated,  $t_h$, as a function of vaccination rate, $\mu$, with $\phi=0.6$. Without or with sufficiently low social measure compliance ($a=0$ and $a=0.5$; red and blue curves), the time until eradication increases with $\mu$. For both cases, vaccination coverage $\phi$ is too low to to eradicate the disease in a completely susceptible population (Figure \ref{fig:Rt1}), and consequently a non-negligible percentage of the population must become infected before $R_t<1$. As $\mu$ increases, the infection dynamics slow down, causing the time it takes for $R_t$ to drop below 1 to increase. For high social measure compliance ($a=0.8$; black curve), the time to eradication decreases with $\mu$. When $a$ is sufficiently large, $\phi=0.6$ is large enough to eradicate the disease through vaccination alone (Figure \ref{fig:Rt1}), and increasing the vaccination rate therefore reduces the time to eradication. Thus, rapid eradication of the virus is only achievable with sustained, widely obeyed social measures.

The right panel of Figure \ref{fig:Rt1mu} shows the peak ICU load over $\mu$. Naturally, the faster the vaccine is administered to the population, the lower the peak ICU case load will be. However, peak ICU load only remains below typical ICU capacity ($1-3\times10^{-4}$) for small $\mu$ when social measure compliance is high (black curve). This again suggests that social measures must remain in place throughout the vaccination program in order to avoid hospital strain and associated loss of life.

\begin{figure}[h!]
{\centering
\includegraphics[width=0.9\textwidth]{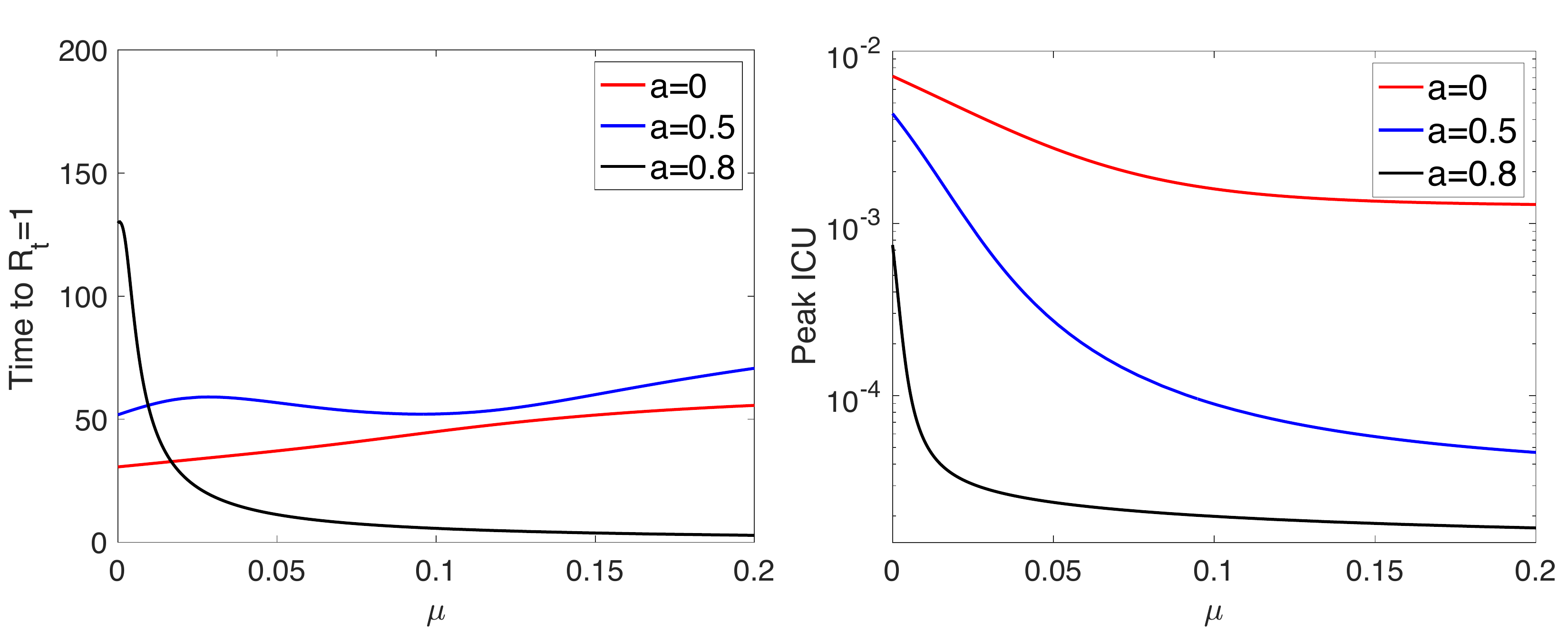}

}
\caption{Controlling the pandemic with a vaccine depends on fast dispersal and sustained social measure compliance. (Left) Time until $R_t<1$ over $\mu$. (Right) Peak ICU cases over $\mu$ (log scale). In both panels, $\phi=0.6$, and the red curve corresponds to no social measures ($a=0$), the blue curve corresponds to moderate social measure compliance ($a=0.5$), and and the black curve corresponds to high compliance ($a=0.8$). Details on how these figures were made are in Appendix \ref{sec:BVPs}.}
\label{fig:Rt1mu}
\end{figure}

%
%

\section{Discussion and Conclusion}

We introduce an extended SEIR socio-epidemiological model incorporating vaccination dynamics to evaluate the interactions between vaccination and social measures for controlling or ending the spread of COVID-19. Following standard analytical techniques \cite{vanDenDriessche2002}, we derived an explicit form for the effective reproduction number $R_t$. This value is of central concern to controlling the pandemic: through a combination of natural infection, social measures, and vaccination administration, we must drive $R_t<1$ in order to eradicate the disease. Our analysis therefore focused on the influence of social measures and vaccination rates on the time until the disease is the eradicated, but also considered hospital demand as a function of these interventions. Importantly, we show that while eliminating social measures entirely might help eradicate the disease faster, the hospital demand, and therefore death toll, are reduced dramatically with even partial adherence to social intervention strategies.

Our analysis focused on three parameters: the proportion of the population willing to receive a vaccine $\phi$, the proportion of the population willing to adhere to social measures $a$, and the rate at which vaccines are administered to the population $\mu$.  Figures \ref{fig:Rt1}-\ref{fig:Rt1mu} summarize the major results arising from these interactions, and suggest that, with low vaccination rate $\mu$, sustained social measures become increasingly important to keep the hospitalization rates low, even if a large proportion of the population are willing to receive the vaccine (that is; even if $\phi$ is large). This finding is consistent with previous studies under varied assumptions \cite{Iboi2020,Brett2020,Choi2020}.

The interplay between social measures and vaccine administration is perhaps most complicated when considering the time until eradication. When the proportion of the population who adhere to social measures is small, the time to eradication is relatively fast (Figures \ref{fig:Rt1phi} and \ref{fig:Rt1mu}, left panels). 
This is because, without social measures, the virus spreads quickly, thereby increasing the number of individuals with infection-conferred immunity (or who die due to the disease). On the other hand, when social measure adherence is high, the time to eradication is large for low vaccination coverage and rate, but decreases with both parameters. For low vaccination coverage or rate, population-level immunity is primarily being conferred via infection, and infection rates are low due to high levels of social measures. As vaccination coverage or rate increases, however, the rate at which individuals become vaccinated begins to outpace the rate at which individuals become infected, and the time until eradication becomes small. Importantly, only in the case of high social measure compliance and high vaccination rate or coverage do the number of ICU cases remain manageable (Figures \ref{fig:Rt1phi} and \ref{fig:Rt1mu}, right panels).

We developed our model under a set of assumptions that captured important features of COVID-19 transmission. However, we omitted one or more realities of COVID-19 dynamics that could quantitatively influence our results. First, we did not include any age structure to our model. It is well known that mortality rates due to COVID-19 are disproportionately high among elderly populations. By including age structure, on can study targeted vaccination programs in which the elderly are given earlier access to the vaccine. Second, we include only a single vaccine in our model, while many with varying efficacies are likely to enter the market before the pandemic is over \cite{Oliver2020A,Oliver2020B,Kaur2020}. While incorporating these features into our model would surely result in quantitative differences, the qualitative predictions of our current model would likely remain unchanged; that is, social measures must remain in place throughout the vaccination campaign in order to mitigate hospital and mortality rates.  An important consideration for future work is the degree to which variants of the SARS-CoV-2 will increase community spread \cite{Mahase2021}, which can be incorporated into our model as a second set of infected classes. 

Notwithstanding these limitations, this work highlights several major implications for the use of vaccination for either controlling or eradicating the current pandemic. The first is that slow vaccination roll out rates mean that continuing with currently applied social measures is imperative to containing the clinical outcomes (demand for ICU care and deaths) of the pandemic for a population. Only a ramped up vaccination rate will allow easing of these social measures. The second important finding arising from the present results is that while social measures under the current slow rate of vaccinations will be crucial to prevent hospitalizations and the death toll from the virus, this intervention will also delay the development of herd immunity in the population. However, two critical results here are that at high levels of social measures, the numbers of individual that are required to be vaccinated to achieve $R_t < 1$ can be significantly small, and that there might be a vaccination coverage level past which this can be achieved. We term this as ``herd immunity due to social measures'', which will be much lower than the corresponding herd immunity in the absence of social measures. Note, however, that the imposition of social measures will keep a large fraction of the population continuing to be susceptible, and while achieving the lower level of herd immunity through vaccination under these measures will allow interruption of transmission, any relaxation of the latter in the presence of infected individuals, or if infected individuals were to arrive into an area lifting such restrictions, would seed resurgences of infection. This conclusion suggests that only by ramping up vaccinations to achieve natural herd immunity (i.e., the higher level of herd immunity that will be required to prevent transmission in the absence of any social containment measures) will the pandemic be fully suppressed over the longer-term.

\vskip1pc

\noindent\textbf{Data accessibility.} All parameter data and code can be found at\\ https://github.com/EdwinMichaelLab/COVID-Vaccination-Paper.}



\section*{Appendix}
\appendix

\section{Time $R_t=1$ and Peak ICU cases}\label{sec:BVPs}

To determine the time until $R_t<1$, and, separately, to determine the peak ICU cases as a function of $\phi$ (Figure \ref{fig:Rt1phi}) and $\mu$ (Figure \ref{fig:Rt1mu}), we consider system \eqref{eq:model} extended to include an auxiliary variable $\tau$:

\begin{equation}\label{eq:BVP}
\begin{aligned}
S'&=-(1-d)\beta S(I_a+I_p+I_m)-\mu S\\
X'&=\mu(1-\phi)S-(1-d)\beta X(I_a+I_p+I_m)\\
V'&=\mu\phi-(1-\xi)(1-d)\beta V (I_a+I_p+I_m)\\
E'&=(1-d)\beta (I_a+I_p+I_m)\left[S+X+(1-\xi) V\right]-\sigma E\\
I_a'&=\sigma\rho E-\gamma_a I_a\\
I_p'&=\sigma(1-\rho)E-\delta_1I_p\\
I_m'&=\delta_1I_p-x_1\delta_2I_m-(1-x_1)\gamma_mI_m\\
I_h'&=x_1\delta_2I_m-x_2\delta_3I_h-(1-x_2)\gamma_hI_h\\
I_c'&=x_2\delta_3I_h-(1-x_3)\gamma_cI_c-x_3mI_c\\
R'&=\gamma_aI_a+(1-x_1)\gamma_mI_m+(1-x_2)\gamma_hI_h+(1-x_3)\gamma_cI_c\\
D'&=x_3mI_c\\
\tau'&=0.
\end{aligned}
\end{equation}
This additional variable provides an additional degree of freedom allowing us to solve the system as a boundary value problem (BVP) with a terminal, parameter-dependent boundary condition (BC). In particular, the variable $\tau$ in solutions of system \eqref{eq:BVP} with BCs
\begin{equation}\label{eq:BCsRt}
\begin{aligned}
S(0)&=S_0\\
E(0)&=1-S_0\\
R_t(\tau)&=1\\
X(0)&=V(0)=I_a(0)=I_p(0)=I_m(0)=I_h(0)=I_c(0)=R(0)=D(0)=0,
\end{aligned}
\end{equation}
where $R_t(t)$ is defined in equation \eqref{eq:RtTime}, represents the time at which $R_t(t)=1$. Since $R_t(t)$ is strictly decreasing over time, $R_t(t)<1$ for all $t>\tau$, and $\tau$ is therefore the time at which the disease is eradicated.

Next, consider system \eqref{eq:BVP} with BCs
\begin{equation}\label{eq:BCsICU}
\begin{aligned}
S(0)&=S_0\\
E(0)&=1-S_0\\
0&=x_2\delta_3I_h(\tau)-(1-x_3)\gamma_cI_c(\tau)-x_3mI_c(\tau)\\
X(0)&=V(0)=I_a(0)=I_p(0)=I_m(0)=I_h(0)=I_c(0)=R(0)=D(0)=0.
\end{aligned}
\end{equation}
The third boundary condition is equivalent to $I_c'(\tau)=0$ and therefore requires $I_c$ to be at a local maximum at time $\tau$. The variable $\tau$ is therefore the time at which $I_c$ reaches its maximum \cite{Young2015}. Evaluating $I_c$ at $t=\tau$ for any solution of BVP \eqref{eq:BVP}-\eqref{eq:BCsICU} provides the peak ICU cases for the fixed parameter set.

Using AUTO \cite{Bard2002}, we can numerically continue solutions to system \eqref{eq:BVP} with BCs \eqref{eq:BCsRt} or \eqref{eq:BCsICU} over any system parameter; doing so over $\phi$ and $\mu$ produced the Figures  \ref{fig:Rt1phi} and  \ref{fig:Rt1mu}, respectively. In both figures, $S_0=0.99$.


\bibliographystyle{RS} 

\bibliography{covidBib} 

\begin{thebibliography}{99}

\bibitem{Randall2021}
Randall T. 2021  When will life return to normal? In 7 years at today's vaccine
  rates..
  https://www.bloomberg.com/news/articles/2021-02-04/when-will-covid-pandemic-end-near-me-vaccine-coverage-calculator.

\bibitem{Lurie2020}
Lurie N, Saville M, Hatchett R, Halton J. 2020  Developing Covid-19 vaccines at
  pandemic speed. {\em New England Journal of Medicine} \textbf{382},
  1969--1973.

\bibitem{Ebrahim2020}
Ebrahim SH, Ahmed QA, Gozzer E, Schlagenhauf P, Memish ZA. 2020  Covid-19 and
  community mitigation strategies in a pandemic. {\em BMJ} \textbf{368}.

\bibitem{Buckner2020}
Buckner JH, Chowell G, Springborn MR. 2020  Dynamic Prioritization of COVID-19
  Vaccines When Social Distancing is Limited for Essential Workers. {\em
  medRxiv}.

\bibitem{Lipsitch2020}
Lipsitch M, Dean NE. 2020  Understanding COVID-19 vaccine efficacy. {\em
  Science} \textbf{370}, 763--765.

\bibitem{Terry2021}
Terry M. 2021  UPDATED Comparing COVID-19 Vaccines: Timelines, Types and
  Prices.
  https://www.biospace.com/article/comparing-covid-19-vaccines-pfizer-biontech-moderna-astrazeneca-oxford-j-and-j-russia-s-sputnik-v/.

\bibitem{Wise2021}
Wise J. 2021  Covid-19: New data on Oxford AstraZeneca vaccine backs 12 week
  dosing interval. {\em BMJ} \textbf{372}.

\bibitem{OurWorld2021}
Ritchie H, Ortiz-Ospina E, Beltekian D, Mathieu E, Hasell J, Macdonald B,
  Giattino C, Roser M. 2021  Coronavirus (COVID-19) Vaccinations - Statistics
  and Research. https://ourworldindata.org/covid-vaccinations.

\bibitem{Davies2020}
Davies NG, Barnard RC, Jarvis CI, Kucharski AJ, Munday J, Pearson CA, Russell
  TW, Tully DC, Abbott S, Gimma A et~al.. 2020  Estimated transmissibility and
  severity of novel SARS-CoV-2 Variant of Concern 202012/01 in England. {\em
  medRxiv}.

\bibitem{Iboi2020}
Iboi EA, Ngonghala CN, Gumel AB. 2020  Will an imperfect vaccine curtail the
  COVID-19 pandemic in the US?. {\em Infectious Disease Modelling} \textbf{5},
  510--524.

\bibitem{Jentsch2020}
Jentsch P, Anand M, Bauch CT. 2020  Prioritising covid-19 vaccination in
  changing social and epidemiological landscapes. {\em medRxiv}.

\bibitem{Newcomb2020}
Newcomb K, Smith ME, Donohue RE, Wyngaard S, Reinking C, Sweet CR, Levine MJ,
  Unnasch TR, Michael E. 2020  Iterative near-term forecasting of the
  transmission and management of SARS-CoV-2/COVID-19 using social interventions
  at the county-level in the United States. .

\bibitem{Young2015b}
Young G, Shim E, Ermentrout GB. 2015  Qualitative effects of monovalent
  vaccination against rotavirus: A comparison of North America and South
  America. {\em Bulletin of mathematical biology} \textbf{77}, 1854--1885.

\bibitem{vanDenDriessche2002}
Van~den Driessche P, Watmough J. 2002  Reproduction numbers and sub-threshold
  endemic equilibria for compartmental models of disease transmission. {\em
  Mathematical biosciences} \textbf{180}, 29--48.

\bibitem{Bard2002}
Ermentrout B. 2002 {\em Simulating, analyzing, and animating dynamical systems:
  a guide to XPPAUT for researchers and students}.
SIAM.

\bibitem{Brett2020}
Brett TS, Rohani P. 2020  Transmission dynamics reveal the impracticality of
  COVID-19 herd immunity strategies. {\em Proceedings of the National Academy
  of Sciences} \textbf{117}, 25897--25903.

\bibitem{Choi2020}
Choi W, Shim E. 2020  Optimal strategies for vaccination and social distancing
  in a game-theoretic epidemiologic model. {\em Journal of theoretical biology}
  \textbf{505}, 110422.

\bibitem{Oliver2020A}
Oliver SE, Gargano JW, Marin M, Wallace M, Curran KG, Chamberland M, McClung N,
  Campos-Outcalt D, Morgan RL, Mbaeyi S et~al.. 2020  The Advisory Committee on
  Immunization Practices’ Interim Recommendation for Use of Pfizer-BioNTech
  COVID-19 Vaccine—United States, December 2020. {\em Morbidity and Mortality
  Weekly Report} \textbf{69}, 1922.

\bibitem{Oliver2020B}
Oliver SE. 2020  The Advisory Committee on Immunization Practices’ Interim
  Recommendation for Use of Moderna COVID-19 Vaccine—United States, December
  2020. {\em MMWR. Morbidity and mortality weekly report} \textbf{69}.

\bibitem{Kaur2020}
Kaur SP, Gupta V. 2020  COVID-19 Vaccine: A comprehensive status report. {\em
  Virus research} p. 198114.

\bibitem{Mahase2021}
Mahase E. 2021  Covid-19: What new variants are emerging and how are they being
  investigated?. {\em BMJ} \textbf{372}.

\bibitem{Young2015}
Young G, Ermentrout B, Rubin JE. 2015  A boundary value approach to
  optimization with an application to salmonella competition. {\em Bulletin of
  mathematical biology} \textbf{77}, 1327--1348.

\end{thebibliography}

\end{document}